\documentclass[twocolumn,amsmath,amssymb,prl,footinbib,floatfix,superscriptaddress]{revtex4}

\usepackage{graphicx}
\usepackage{amsmath}
\usepackage{amsfonts}
\usepackage{color}
\usepackage{slashed}
\usepackage{stackrel}

\font\numbers=cmss12
\font\upright=cmu10 scaled\magstep1
\def\stroke{\vrule height8pt width0.4pt depth-0.1pt}
\def\topfleck{\vrule height8pt width0.5pt depth-5.9pt}
\def\botfleck{\vrule height2pt width0.5pt depth0.1pt}
\def\Zmath{\vcenter{\hbox{\numbers\rlap{\rlap{Z}\kern
0.8pt\topfleck}\kern 2.2pt
                   \rlap Z\kern 6pt\botfleck\kern 1pt}}}
\def\Qmath{\vcenter{\hbox{\upright\rlap{\rlap{Q}\kern
                   3.8pt\stroke}\phantom{Q}}}}
\def\Nmath{\vcenter{\hbox{\upright\rlap{I}\kern 1.7pt N}}}
\def\Rmath{\vcenter{\hbox{\upright\rlap{I}\kern 1.5pt R}}}
\def\Pmath{\vcenter{\hbox{\upright\rlap{I}\kern 1.5pt P}}}

\def\barray{\begin{eqnarray}}
\def\earray{\end{eqnarray}}
\def\beq{\begin{equation}}
\def\eeq{\end{equation}}

\begin{document}

\date{\today}

\title{Platonic Entanglement}
\author{Jos\'e I. Latorre$^{1,2,3}$ and Germ\'an  Sierra$^4$ \vspace{0.2cm}
 \\  
${}^1$  Quantum Research Centre, Technology Innovation Institute, Abu Dhabi, UAE. \\
${}^2$ Center for Quantum Technologies, National University of Singapore, Singapore. \\
${}^3$ Qilimanjaro Quantum Tech, Barcelona, Spain  \\ 
 ${}^4$  Instituto de F\'isica Te\'orica UAM-CSIC, Universidad Aut\'onoma de Madrid, Cantoblanco, 
 Madrid, Spain.}

\begin{abstract}
We present a construction of highly entangled states defined on the topology of a platonic solid using tensor networks based on ancillary Absolute Maximally Entangled (AME) states. We illustrate the idea using the example of a quantum state based on AME(5,2) over a dodecahedron.  We analyze the entropy of such states on many different partitions, and observe that they come on integer numbers and are almost maximal.
We also observe that all platonic solids accept the construction of AME states based on Reed-Solomon codes since their number of facets, vertices and edges are always a prime number plus one.

\end{abstract}

\maketitle

\bigskip
\section{Entanglement and Tensor Networks}

The way multipartite entanglement is distributed over a quantum system is key to understand its emerging properties and, also, to devise good classical approximation strategies. A successful technology to achieve this dual goal is to represent quantum states as Tensor Networks (TN) \cite{TN,TN2}. Indeed, 
TN provide one of the most powerful methods to represent classically a quantum state as a contraction of a series of tensors that take indices in an ancillary space. 

It can be argued that TN attempt to
describe a quantum state on a basis, different from the computational one, that rightly fits the entanglement properties of the system. Therefore, the technology of TN  can be considered as a bona fide method to exploit the actual Kolmogorov complexity of a quantum state. They provide a most economical and adaptive description of quantum correlations.

Symmetries play a relevant role when choosing the appropriate TN to describe a quantum state. Most of condensed matter systems are defined on
regular lattices, favoring the use of variants of TN such as Matrix Product States (MPS),  or Projected Entangled Pair States (PEPS). 
At criticality, scale invariance is better captured  by Multiscale Entanglement Renormalization Ansatz (MERA). 
In general, TN should adapt to the geometry dictated by the Hamiltonian of the system to implement the area law of the entanglement entropy. 
Although there are exceptions to this rule such as the rainbow state \cite{rain1,rain2}. 

It is perfectly correct to try TN which are not based on entangling only pairs of ancillary degrees of freedom. For instance, it may be advantageous 
to use TN  based on three-body entangled units to describe triangular lattices. It can be argued that standard pair-wise entangling units, as in MPS and PEPs, will describe any entangling structure given a sufficient large ancillary dimension. This is indeed true, thought such an approach may not be the most efficient one.
Some efforts have been devoted to analyze the properties of TN based on GHZ and W states to analyze frustrated systems \cite{AL13}. 
Multipartite ancillary states have also been used to represent solutions to 3SAT problems \cite{GL12} 

A more sophisticated use of TN   has been explored in the context of holography.
There, TN  based on Absolute Maximally Entangled (AME) \cite{AME1}-\cite{AME5} states are used to understand the emergence of space time and its holography properties \cite{LS15,PYHP15,H16,HN16,J21}.

Here, we consider the playful idea of building quantum states as the result of using AME states on topologies dictated by platonic solids. It is not obvious how such states can be built and, if so, how entanglement will be globally distributed. We shall find that large entanglement gets distributed if TN  based on AMEs are used. 

Let us also mention that the idea to choose the directions of quantum measurements according to the vertices of Platonic solids has been pursued in Ref \cite{G20}. Other works where Platonic solids have appeared in the context of Quantum Information can be found in references \cite{other1}-\cite{other5}.


\section{Absolute Maximally Entangled States}

Absolute Maximally Entangled \cite{AME1}-\cite{AME5} states, also called Perfect states \cite{PYHP15}, are defined as those multi-partite quantum 
states whose reduced density matrices for any bi-partition are proportional
to the identity. To be precise, given a state of $n$ qudits, 
$|\psi \rangle  \in {\cal H}=(\mathbb{C}^d)^{\otimes n}$, this state will be an AME state 
if all its reduced density matrices to a part {\cal A} of $m$ degrees of freedom, such that ${\cal A} \otimes \bar {\cal A}={\cal H}$,
carries a von Neumann entropy $S_{\cal A} =-{\rm tr} (  \rho_{\cal A} \log_2 \rho_{\cal A})$ given by
\begin{equation}
  S_{\cal A}=m \log_2 d  \, . 
\end{equation}
This is equivalent to finding that all the reduced density matrices for $m\le n/2$ are proportional to the identity
\begin{equation}
 \rho_m=\frac{1}{d^m}I_{d^m}  \,  .
\end{equation}
We shall be considering states made with an even number of parts.
It is then sufficient to verify that even bi-partitions are maximally entangled to guarantee that the same properties holds for smaller partitions.

 Absolutely Maximally Entangled states are usually labeled
as AME($n$,$d$), where  $n$ is  the number of local
degrees of freedom and  $d$ is the local dimension. 
In general, there is an obstruction to find AME states; for certain values of $n$ and $d$ they may not exist any. 
For instance, there is no 4-qubit maximally entangled state in all its partitions, that is,  there is no AME(4,2). 
However, there exist  AME(5,2) and AME(6,2) states. It is proven that there are no  AME states for $n>6$ and $d=2$
\cite{S04,H17}, or for $n=7$ and $d=5$ \cite{B17,B18} (see also \cite{HE18} for $d=3,4,5$ and several $n$'s). 
Some  applications  of AME states is to test the efficiency  to implement multipartite quantum protocols in quantum computers  \cite{A19}
and as  quantum repeaters \cite{A21}.

Let us give an explicit form of AME(5,2) 
\begin{equation}
\label{3}
| {\rm AME}(5,2)\rangle= \frac{1}{\sqrt {2^5}}\sum_{i=0}^{2^5-1} c^{(5,2)}_i |i\rangle,
\end{equation}
where we used the usual shorthand notation for the elements in the computational basis and the coefficients 
have the same modulus and signs given by 
\cite{AME1}-\cite{AME4} 
\begin{eqnarray}
c^{(5,2)}&=&\{1, 1, 1, 1, 1, -1,-1, 1, 1, -1,-1,\nonumber\\ 
&&1, 1, 1, 1,1,1, 1, -1, -1,1, -1,1,\nonumber\\ 
&&-1, -1, 1, -1, 1, -1, -1, 1, 1\} .
\label{ame52}
\end{eqnarray}

A more explicit form is given in Table I, which shows that this state
is not invariant under rotations of the qubit positions, say
$|s_1 s_2 s_3 s_4 s_5 \rangle \rightarrow |s_5 s_1 s_2 s_3 s_4 \rangle$.
We shall give  below another AME(5,2)  state with this property.

\begin{table}[h!]
\begin{tabular}{|r|ccccc|c|c|c|ccccc|c|}
  \hline
$i$ & $s_1$ &  $s_2$ & $s_3$ &  $s_4$ &  $s_5$ & $c_i$ & & 
$i$ & $s_1$ &  $s_2$ & $s_3$ &  $s_4$ &  $s_5$ & $c_i$  \\
\hline
0 & 0 & 0 & 0 & 0 & 0 & + &  & 16 & 1 & 0 & 0 & 0 & 0 & + \\
1 & 0 & 0 & 0 & 0 & 1 & + &  & 17 & 1 & 0 & 0 & 0 & 1 & + \\
2 & 0 & 0 & 0 & 1 & 0 & + &  & 18 & 1 & 0 & 0 & 1 & 0 & - \\
3 & 0 & 0 & 0 & 1 & 1 & + &  & 19 & 1 & 0 & 0 & 1 & 1 & - \\
4 & 0 & 0 & 1 & 0 & 0 & + &  & 20 & 1 & 0 & 1& 0 & 0 & + \\
5 & 0 & 0 & 1 & 0 & 1 & - &  & 21 & 1 & 0 & 1 & 0 & 1 & - \\
6 & 0 & 0 & 1 & 1 & 0 & - &  & 22 & 1 & 0 & 1 & 1 & 0 & + \\
7 & 0 & 0 & 1 & 1 & 1& + &  & 23 & 1 & 0 & 1 & 1 & 1 & - \\
8 & 0 & 1 & 0 & 0 & 0 & + &  & 24 & 1 & 1 & 0 & 0 & 0 & - \\
9 & 0 & 1 & 0 & 0 & 1 & - &  & 25 & 1 & 1 & 0 & 0 & 1 & + \\
10 & 0 & 1 & 0 & 1 & 0 & - &  & 26 & 1 & 1 & 0 & 1 & 0 & - \\
11 & 0 & 1 & 0 & 1 & 1 & + &  & 27 & 1 & 1 & 0 & 1 & 1 & + \\
12 & 0 & 1 & 1 & 0 & 0 & + &  & 28 & 1 & 1 & 1 & 0 & 0 & - \\
13 & 0 & 1 & 1 & 0 & 1 & + &  & 29 & 1 & 1 & 1 & 0 & 1 & - \\
14 & 0 & 1 & 1 & 1 & 0 & + &  & 30 & 1 & 1 & 1 & 1 & 0 & + \\
15 & 0 & 1 & 1 & 1 & 1 & + &  & 31 & 1 & 1 & 1 & 1 & 1 & + \\
\hline
\end{tabular}
  \caption{The state \eqref{ame52} with $\pm$ denoting the values $\pm 1$}.

\end{table}

The state \eqref{3}  can be cast in different forms using local unitaries, that will never change its entanglement properties.

Let us also provide the absolutely maximal entangled 6-qubit state
\begin{equation}
| {\rm AME}(6,2)\rangle= \frac{1}{\sqrt {2^6}}\sum_{i=0}^{2^6-1} c^{(6,2)}_i |i\rangle,
\end{equation}
with 
\begin{eqnarray}
c^{(6,2)}&=&\{
 -1,- 1,- 1,+ 1,- 1, 1, 1, 1, \nonumber\\ 
&&-1,- 1,- 1, 1,1,- 1,- 1,- 1, \nonumber\\ 
&&-1, -1, 1, -1,- 1, 1, -1,- 1, \nonumber\\ 
&&1, 1, -1, 1,- 1, 1, -1,- 1, \nonumber\\ 
&&-1, 1, -1,- 1,- 1,- 1, 1,- 1, \nonumber\\ 
&&1, -1, 1,1,- 1,- 1, 1,- 1, \nonumber\\ 
&&1, -1,- 1,-1, 1, 1, 1,- 1, \nonumber\\ 
&&1, -1,- 1,-1,- 1,- 1,- 1, 1 \} .
\label{ame62}
\end{eqnarray}

A less trivial example of AME state corresponds to the case of 4 qutrits. Its explicit form is
\begin{equation}
|{\rm AME}(4,3)\rangle =  \frac{1}{9} \sum_{i,j,=0,1,2} |i\rangle |j\rangle |i+j\rangle |i+2 j\rangle \ ,
\end{equation}
where all labels are understood as mod 3. Each bi-partition in 2 + 2 qutrits 
 of the above state carries a von Neumann entropy $S=2 \log_2 3$. 
 This state was used in \cite{LS15} to construct an  holographic code
 where 3 qutrits were ancillae  and one qutrit was a physical variable.

\section{Platonic States}

We shall now construct novel quantum states which are constructed as
a contraction of TN   defined on the geometry of a platonic solid.
We call these states, Platonic states.

\subsection{Qubits on vertices of a dodecahedron}

\begin{figure}[ht]
\centering
\includegraphics[width=0.25\textwidth]{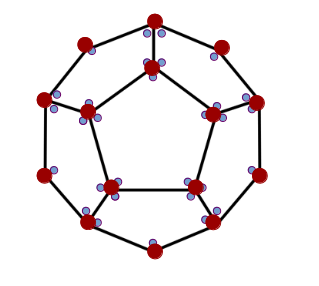}
\caption{A quantum state is defined on the vertices of a dodecahedron using an underlying tensor network. Each qubit is colored in red, whereas the ancillary degrees of freedom associated to 5-qubit maximally entangled states   live on the pentagon faces and are represented in violet.
 } 
\label{dodecahedron} 
\end{figure}

To illustrate the construction of a platonic state, we consider the explicit example
of a quantum state defined on the 20 vertices of a dodecahedron
(see Fig. \ref{dodecahedron}). The way to construct this state consists on producing a tensor network similar in spirit to the well-known PEPs, but using 5-qubit maximally entangled states instead of the usual maximally entangled pairs. The intuition behind this construction is that the AME(5,2) state may be able to distribute entanglement in a very efficient manner.

\begin{figure}[ht]
\centering
\includegraphics[width=0.25\textwidth]{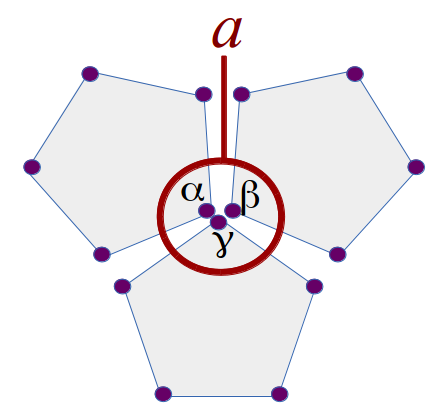}
\caption{Basic tensor assignment. Three pentagons meet at a vertex, each carrying
 ancillary indices $\alpha$, $\beta$ and $\gamma$, respectively. The tensor gets defined as $A^a_{\alpha\beta\gamma}=\delta^a_\alpha \delta^a_\beta \delta^a_\gamma$. Thus, the physical index $a$ takes the value of the ancillae when they coincide, otherwise it is set to $0$.
 } 
\label{simplex} 
\end{figure}

To be precise, the construction starts by filling the 12 pentagons of the dodecahedron with AME(5,2). Then, we define physical indices in each vertex using an agreement clause (see Fig. \ref{simplex})
\beq
A^a_{\alpha\beta\gamma}=\delta^a_\alpha \delta^a_\beta \delta^a_\gamma .
\label{tensor}
\eeq
That is, physical indices are made to coincide with the ancillary ones when all of them agree.

Since the basic tensor defining AME(5,2) in Eq. (\ref{ame52})  is made of $\pm 1$, 
the  global state we are defining on a dodecahedron will carry all possible superpositions
weighted with a plus or minus sign
\begin{equation}
| D_1 \rangle= \frac{1}{\sqrt {2^{20}}}\sum_{i=0}^{2^{20}-1} c^{D}_i |i\rangle,
\label{D1} 
\end{equation}
that is, all $c^{D}_i= 1$\ or\ $-1$.

The entanglement properties of the emerging state on a dodecahedron are non-trivial.
The von Neumann entropy $S_A(D_1)$ for the state \eqref{ame52} and several block sizes $|A|$ 
 are collected in Table II. We first observe that all bi-partitions, local or non-local, that involve  6 qubits or less carry maximal entanglement.
For the bi-partitions of 10 vs 10 spins, the result for the entropy turns out to be always an integer number, and its maximal value of 10 is attainable. 

\begin{table}[h!]
\begin{tabular}{|r||c|c|c|c|c|c|c|c|c|c|}
\hline
$|A|$ & 10 & 9 & 8 & 7 & 6 & 5 & 4 & 3 & 2 & 1 \\
\hline
$S_A(D_1)$ & 7,8,9,10 & 7,8,9 & 7,8 & 6,7 & 6 & 5 &  4 &  3 &  2 &  1 \\
\hline 
$S_A(D_2)$ & 7,8  & 7,8& 6,7,8 & 6,7 & 5,6 & 4,5 &  3,4 &  3 &  2 &  1 \\
\hline 
\end{tabular}
  \caption{Entanglement entropies of AME(5,2) states \eqref{D1}  for various blocks of sizes $|A|$. 
  $S(D_1)$ are for the state based on \eqref{ame52} and  $S(D_2)$ are for the state using  \eqref{am1}.  }.

\end{table}

\subsection{A rotational invariant AME(5,2) state}

\begin{figure}[ht]
\centering
\includegraphics[width=0.6\textwidth]{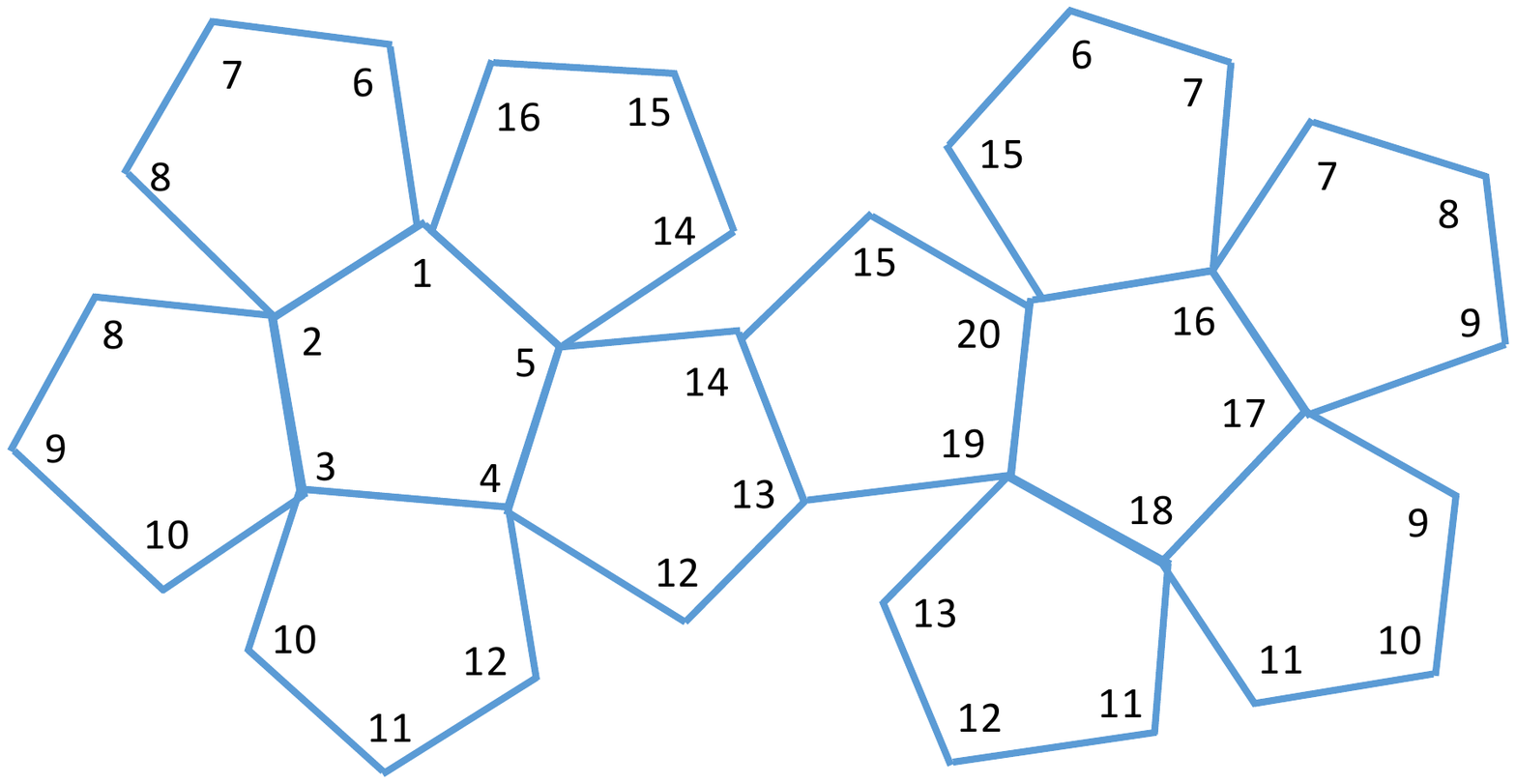}
\vspace{-2.5cm} 
\caption{ Labels of the vertices of the dodecahedrum.
 } 
\label{dodecahedro} 
\end{figure}

We observed above that the state \eqref{ame52} is not invariant under rotations
of the qubits around  the vertices of the pentagon. This implies that the
resulting state over  the dodecahedron depends on the specific choice of the AME(5,2) state
for every pentagon. The analogy is that of a dodecahedron made of pentagons with different colors.
The counting number of emerging  states is an interesting but difficult problem that we shall not deal with here. 
Let us  consider a rotational invariant  AME(5,2) state given by \cite{AME1}-\cite{AME4}
\barray 
&&|{\rm AME}(5,2) \rangle =  \frac{1}{4}    \left[ 
|00000 \rangle + |10010 \rangle + |01001 \rangle
\right. 
\label{newame52} \label{am1}    \\ & & 
 +  | 10 100 \rangle + |0 1010 \rangle + |00101\rangle - |01111 \rangle - |10111 \rangle 
\nonumber  \\ & & 
- |11011 \rangle - |11101 \rangle - |11110 \rangle - |11000 \rangle- |01100  \rangle  
\nonumber \\
& & \left.  - | 00110 \rangle - |00011 \rangle - |10001 \rangle \right]  \, . 
\nonumber
\earray 
This state can be written as
\barray 
|{\rm AME}(5,2) \rangle & =  &  \frac{1}{4}   \sum_{ \sum_j  s_j =0 \; ({\rm mod} \, 2)}  (-1)^{ \sum_j s_j s_{j+1} }    \label{amebis} \\ 
 &  &  \qquad  \qquad   \qquad  \times |s_1, \dots, s_5 \rangle  \ , 
\nonumber 
\earray 
where the sum runs  over the spin  configurations  whose addition  vanishes  mod 2  (we take $s_6 = s_1$). 
Let us  write (\ref{amebis}) as 
\beq
|P \rangle = \frac{1}{4} \sum_{s_j , j\in P} (-1)^{ \eta_P}   |s_1, \dots, s_5 \rangle , \quad \eta_P =  \sum_{j \in P} s_j s_{j+1} \ , 
\label{P}
\eeq
where $P$ denotes the  pentagon whose vertices are occupied by the spins $s_1, \dots, s_5$. 
The state over the dodecahedron constructed using (\ref{tensor}) and  (\ref{P}),   reads  
\beq
|D_2   \rangle  = a_D   \sum_{s_1, \dots, s_{20}}  (-1)^{  \eta_{P_1} + \dots +  \eta_{P_{12}}}  |s_1, \dots, s_{20} \rangle ,
\label{D-state} 
\eeq
where the spins on each pentagon satisfy the neutrality condition 
\beq
\sum_{i \in P_a} s_i  = 0 \;  ( {\rm mod} \, 2 ),  \; \;   a=1, \dots, 12  \, . 
\label{cons}
\eeq
$a_D$ is the  normalization constant of the state that is derived below.  
The  asignement  of vertices  $i=1, \dots, 20$,   to pentagons 
$P_a \; (a=1, \dots, 12$) is  given in Fig. \ref{dodecahedro}, that 
yields the neutrality conditions : 
$0 = s_1 + s_2 + s_3 + s_4 + s_5  =   s_6 + s_7 + s_{15}  + s_{16}  + s_{20}  = \dots$
where the equalities are defined mod  2. 
%
%
This is  a set of  12  linearly independent equations for 20 variables 
that leaves   8 undetermined spins.  
Hence the state (\ref{D-state}) is the linear superposition of  $2^8$ states out of the $2^{20}$ states of  the computational
basis that is normalized with  $a_D = 1/\sqrt{ 2^{8}}$.  Another feature of 
(\ref{D-state}) is that the sign factor $(-1)^{\sum_a \eta_{P_a}}$ is equal to 1 for the $2^8$ 
spin configurations that satisfy the neutrality conditions (\ref{cons}). 
Indeed,  let's take a pair of spins, say $s_i$ and  $s_j$, on a link $\langle i, j \rangle$
on the dodecahedrum.   They  give   
a factor   $(-1)^{s_i s_j}$,  associated to  the pentagons sharing  the link $\langle i, j \rangle$, and so  the total contribution is  1.
Hence the state (\ref{D-state})  takes finally   the form
\beq
|D_2  \rangle  = \frac{1}{\sqrt{2^8}}    \sideset{}{'}  \sum_{s_1, \dots, s_{20}}    |s_1, \dots, s_{20} \rangle \, , 
\label{D-state2}
\eeq
where $\sideset{}{'}  \sum$ denotes the  sum  over the spin configurations that satisfy (\ref{cons}).

The entanglement entropies $S_A(D_2)$  of (\ref{D-state2}) are collected 
in Table II for several blocks. Their maximal  values are always  lower than  those of  the state \eqref{D1}.
An heuristic explanation is that \eqref{D-state2} contains half of the states of \eqref{D1} of the computational basis. 

The entanglement entropies of \eqref{D-state2}  can be easily  computed.
Let $A$ be a subset of  $V_A  \leq 10$ vertices. In general, not all the corresponding  spins 
  will be independent variables. 
For example,  if $A$ contains  the sites $i=1,2,3,4,5$  forming  a pentagon, their spins must  satisfy that 
$\sum_{i=1}^5 s_i = 0 \; ({\rm mod} \;  2)$, hence  only 4 spins are independent.   In this case the state (\ref{D-state2})  reads
\beq
|D \rangle  = \frac{1}{\sqrt{2^4}}    \sideset{}{'}  \sum_{s_1, \dots, s_{5}}    |s_1, \dots, s_{5} \rangle \otimes  
\frac{1}{\sqrt{2^4}}    \sideset{}{'}  \sum_{s_6, \dots, s_{20}}    |s_6, \dots, s_{20} \rangle
 \, , 
\label{D-state3}
\eeq
yielding  an entropy $S_A = 4$.  If the subset $A$
contains $n_A$ independent  spin variables then (\ref{D-state2}) is  decomposable  as 
\beq
|D \rangle  = \frac{1}{\sqrt{2^{n_A}}}    \sideset{}{'}  \sum_{s_A}    |s_A \rangle \otimes  
\frac{1}{\sqrt{2^{8-n_A}}}    \sideset{}{'}  \sum_{s_B}    |s_B \rangle
 \, , 
\label{D-state4}
\eeq
that gives an  entropy $S_A = n_A$. 
Obviously, $n_A \leq 8$, which is the maximal attainable  entropy. This can be achieved for 
$A= \{ 1,2,3,4,5,16,17,18,19,20 \}$ that contains two 
pentagons on opposite sides of  the dodecahedrum  (see Fig.\ref{dodecahedro}).

\subsection{Hovering qubits}

A different arrangement of qubits based on AME networks can be 
obtained as follows. Every one of the 12 pentagons
in the dodecahedron will carry an associated
extra "hovering" qubit. We may think of it as placed in the center of the pentagon.
This defines a unit cell made out of 6 qubits,  five of them  in the pentagon plus the
hovering one. On each of these cells we introduce an AME(6,2).

We can then analyze the bipartitions of the 12-qubit state, either local or non-local.
The resulting reduced density matrices to 6 qubits carry entropies
that range from 4 to 6.

\section{Reed-Solomon codes for each platonic solid}

Reed-Solomon codes offer the possibility of constructing fully AME states \cite{RS60}.
It is though clear that  such codes do not make any use of the specific geometry
of the dodecahedron. They just depend on the number of $d$-dimensional degrees
of freedom placed  on $n$ sites. In particular, Reed-Solomon codes exist for 
states with $d=p$ a  prime number,  and $n=p+1$. This fact combines smoothly with
the 5 platonic solids, since in all cases, the number of vertices, edges  and faces are 
 a prime number plus one. 

\begin{table}
\begin{tabular}{|l|l|l|}
  \hline
  Tetrahedron& Faces = 4 & AME(4,3)\\
  Tetrahedron& Edges = 6 & AME(6,5)\\
  Tetrahedron& Vertices = 4 & AME(4,3)\\
 \hline
  Exahedron& Faces = 6 & AME(6,5)\\
  Exahedron& Edges = 12 & AME(12,11)\\
  Exahedron& Vertices = 8 & AME(8,7)\\
 \hline
  Octahedron& Faces = 8 & AME(8,7)\\
  Octahedron& Edges = 12 & AME(12,11)\\
  Octahedron& Vertices = 6 & AME(6,5)\\
 \hline
  Dodecahedron& Faces = 12 & AME(12,11)\\
  Dodecahedron& Edges = 30 & AME(30,29)\\
  Dodecahedron& Vertices = 20 & AME(20,19)\\
 \hline
  Icosahedron& Faces = 20 & AME(20,19)\\
  Icosahedron& Edges = 30 & AME(30,29)\\
  Icosahedron& Vertices = 12 & AME(12,11)\\

  \hline
\end{tabular}
  \caption{All platonic solids carry a number of Faces, Edges and Vertices which are a prime number plus one. 
  Thus, a Reed-Solomon code can be associated to each construction.}

\end{table}

Let us illustrate the example of AME(12,11), such that the minimal Hamming distance between any pair of elements in the
set is $d_H=7$.
We first need to create the Reed-Solomon generating matrix, made with increasing powers of integers 1 to 10, all ${\rm mod}(11)$,
\begin{equation}
G= \left( 
 \begin{array}{cccccccccccc}
1&1&1&1&1&1&1&1&1&1&1&0\\
0&1&2&3&4&5&6&7&8&9&10&0\\
0&1&4&9&5&3&3&5&9&4&1&0\\
0&1&8&5&9&4&7&2&6&3&10&0\\
0&1&5&4&3&9&9&3&4&5&1&0\\
0&1&10&1&1&1&10&10&10&1&10&1
\end{array}  
\right) .
\end{equation}
We then create all the elements of the superposition of AME(12,11)
by taking each element of a basis $x_i$ for 6 11-dits and compute $a_i=x_i\cdot G$. These numbers
become the coefficients of the AME state,
\begin{equation}
|{\rm AME}(12,11)\rangle = \frac{1}{\sqrt{11^6}} \sum_{i=1,\ldots,11^6}| a_i\rangle .
\end{equation}
The result is a state which is maximally entangled in all its partitions, and
with a minimum Hamming distance 7 among each pair of superposed elements.

 In general, the ${\rm AME}(n,p)$ with $n=p+1$ is obtained following a similar procedure. The elements of the superposition are obtained as the images of all elements of $n/2$ bits. The minimal Hamming distance is then $\frac{n}{2} +1$.

\section{Conclusion}

TN based on multipartite entangled ancillary states can be used to deploy entanglement on different topologies. We have here shown how these states define very highly entangled states for the case of using AME ancillary states on platonic solids. The quantum states constructed in this manner are locally fully entangled and achieved very large entanglement on arbitrary partitions. 

It is a remarkable fact that all platonic solids have a number of faces,  edges and vertices corresponding to a prime number plus one. It follows that they all accept AME related to Reed-Solomon error correcting codes as the ancillary building blocks of the TN structure.

\bigskip \bigskip

{\sl Acknowledgements.}

 GS acknowledges 
support from the grants PGC2018-095862-B-C21, 
QUITEMAD+ S2013/ICE-2801, SEV-2016-0597 of the
{\em Centro de Excelencia Severo Ochoa}   Programme and the
CSIC Research Platform on Quantum Technologies PTI-001.

\end{document}